\documentclass[twocolumn,showpacs,%
  aps,superscriptaddress,%
  eqsecnum,prd,notitlepage,showkeys,
  footinbib, floatfix]{revtex4-1}

\usepackage{amssymb}
\usepackage{amsmath}
\usepackage{graphicx}
\usepackage{dcolumn}
\usepackage[colorlinks,urlcolor=blue,citecolor=blue,linkcolor=blue]{hyperref}
\usepackage{lipsum}
\usepackage{color,units}
\usepackage[dvipsnames]{xcolor} 
\usepackage{aas_macros}
\usepackage{lineno}
\usepackage{xspace}

\usepackage{subfigure}

\usepackage{siunitx}
\DeclareSIUnit \parsec {pc}

\usepackage[normalem]{ulem} 

\newcommand{\bilby}{\textsc{bilby}\xspace}

\newcommand{\nrsur}{\textsc{NRSur7dq4}\xspace}
\newcommand{\nrhybsur}{\textsc{NRHybSur3dq8}\xspace}

\newcommand{\imrx}{\textsc{IMRPhenomXHM}\xspace}

\newcommand{\BF}{\mathrm{BF}}

\newcommand{\SPA}{School of Physics and Astronomy, Monash University, Clayton, VIC 3800, Australia}
\newcommand{\OzGravMonash}{OzGrav: The ARC Centre of Excellence for Gravitational Wave Discovery, Clayton, VIC 3800, Australia}

\begin{document}

\title{The Memory Remains (Undetected): Updates from the Second LIGO/Virgo Gravitational-Wave Transient Catalog}

\author{Moritz H\"ubner}
    \email{email@moritz-huebner.de}
\author{Paul Lasky}
    \email{paul.lasky@monash.edu}
\author{Eric Thrane}
    \email{eric.thrane@monash.edu}

\affiliation{\SPA}
\affiliation{\OzGravMonash}

\begin{abstract}
The LIGO and Virgo observatories have reported 39 new gravitational-wave detections during the first part of the third observation run, bringing the total to 50.
Most of these new detections are consistent with binary black-hole coalescences, making them suitable targets to search for gravitational-wave memory, a non-linear effect of general relativity.
We extend a method developed in previous publications to analyse these events to determine a Bayes factor comparing the memory hypothesis to the no-memory hypothesis. 
Specifically, we calculate Bayes factors using two waveform models with higher-order modes that allow us to analyse events with extreme mass ratios and precessing spins, both of which have not been possible before.
Depending on the waveform model we find a combined $\ln \mathrm{BF}_{\mathrm{mem}} = 0.024$ or $\ln \mathrm{BF}_{\mathrm{mem}} = 0.049$  in favour of memory.
This result is consistent with recent predictions that indicate $\mathcal{O}(2000)$ binary black-hole detections will be required to confidently establish the presence or absence of memory.
\end{abstract}

\maketitle


\section{Introduction}\label{sec:introduction}
The gravitational-wave memory effect is a non-oscillatory part of any gravitational wave.
It can be understood as the part of the gravitational wave that is sourced by previously emitted waves.
Memory causes a permanent distortion in spacetime long after the wave has passed~\cite{ZeldovichYa.1974, Braginsky1987, DemetriosChristodoulou1991, Thorne1992, Blanchet1992}.
Memory effects are not included in most numerical relativity models and are hence typically not incorporated in gravitational waveforms of compact binary coalescences.
This is because memory appears in in the $m=0$ modes of the waveform which are difficult to resolve with numerical relativity simulations~\cite{Favata2010,Pollney2011, Mitman2020}.

The slow build-up of memory during compact binary coalescences causes low-frequency contributions to the gravitational-wave signal.
Due to the relatively low sensitivity at lower frequencies ($f \lesssim \SI{20}{\hertz}$), measuring memory in any individual compact binary detection with current generation gravitational-wave observatories is extremely difficult~\cite{Lasky2016, Johnson2018}.
However, recent advances in modelling and gravitational-wave signal analysis have made it possible to coherently search for the presence of memory in an ensemble of gravitational-wave signals~\cite{2018Bilby, BilbyGWTC1, ThanksForTheMemory, Talbot2018, Lasky2016}. 
Previous studies have shown that thousands of gravitational-wave detections with LIGO/Virgo~\cite{aLigo_2010, aLigo_2015, aLigo_GW150914, aLigo_quantum, aLigo_sensitivity_O1O2, aLIGO_sensitivity_O3, aLIGOVirgoKagraProspects, aVirgo, aVirgo_squeeze, GWTC1, GWTC2} operating at their design sensitivity may be required to confidently detect memory, a milestone which is likely to occur during the LIGO A+/Virgo+ era~\cite{ThanksForTheMemory, BoersmaMemory}.

Since memory arises due to explicitly non-linear effects in general relativity, any measurement of memory can be considered to be a test of the theory in this regime.
In this study, we focus on measuring the presence of memory in the gravitational-wave signal itself.
Beyond the detection of memory, future studies will then be able to focus on tests of the exact amplitude and shape of the memory part of the wave, which are motivated by modified theories of gravity~\cite{Garfinkle2017, Hollands2017, Satishchandran2018, Yang2018}, cosmology~\cite{Bieri2016, Bieri2017} and possibly cross-checks with waveform models~\cite{Khera2021}.



\section{Methods}\label{sec:methods}
We follow mostly the description laid out in Ref. \cite{ThanksForTheMemory}, but make some adjustments to our waveform models.
For a detailed description of our methods see Refs.~\cite{Lasky2016, ThanksForTheMemory}.
Recent advances in waveform modelling have made it more feasible to perform lengthy sampling processes with models that include higher-order modes (see e.g. Refs.~\cite{IMRPhenomXHM, IMRPhenomXPHM, PhenomXStudy, NRHYBSUR, Varma2019}).
Including higher-order mode effects into the analysis is not just necessary to avoid systematic errors in the waveform models that could affect the analysis, they are also required to break a degeneracy that leaves the sign of the memory ambiguous~\cite{Lasky2016}.
This degeneracy arises because memory changes its sign under a simultaneous $\SI{90}{\degree}$ rotation of the polarisation angle $\psi$ and the phase at coalescence $\phi_c$.
The same transformation leaves the quadrupolar (leading order) modes unchanged.

For our analysis we use two waveform models.
First, the \imrx waveform model~\cite{IMRPhenomXAS, IMRPhenomXHM, PhenomXStudy}, which is an aligned-spin model and includes several of the most dominant higher-order modes.
Aligned-spins in this context refer to the black-hole spins being parallel (or antiparallel) to the orbital angular momentum of the binary.
\imrx covers mass ranges of the more extraordinary gravitational-wave observations such as GW190814~\cite{GW190814}, which has not been possible for memory analyses until now~\cite{ThanksForTheMemory}.
Furthermore, \imrx natively provides the time domain representation of the waveform in the spherical harmonic mode decomposition, which is required to calculate the memory contribution using the \textsc{gwmemory} package~\cite{Talbot2018}.
Finally, \imrx is among the most computationally efficient waveform models that contain higher-order modes~\cite{IMRPhenomXHM}.
Additionally, we perform our analysis with the \nrsur waveform model~\cite{Varma2019}.
While \nrsur only extends to a mass ratio $q=m_2/m_1=1/4$, which mainly excludes GW190412, GW190814, and parts of the posterior distributions of a few other events, it does include spin precession effects, which have been shown to exist in the population of binaries~\cite{LIGO_RATES_AND_POPULATIONS_O3a}.
\nrsur has been trained on numerical-relativity simulations to a mass ratio of $q=1/4$, and has been shown to work well down to $q=1/6$~\cite{Varma2019}. 
We conservatively restrict the minimum of the prior to $q=1/4$.
Using both \imrx and \nrsur provides us with a cross-check on our results and allows us to deploy at least one model on all events~\footnote{There are more waveform models such as \textsc{IMRPhenomXPHM} that both model extreme mass ratios and precession~\cite{IMRPhenomXPHM}, however it currently poses practical issues for memory calculations.
The \textsc{IMPPhenomXPHM} implementation in the \textsc{LALSuite} package~\cite{LALSuite} only provides the spherical-harmonic mode decomposition in the frequency domain.
Frequency-domain waveforms within \textsc{LALSuite}, unlike time-domain waveforms, are defined so that the merger is exactly at the beginning/end of the data segment.
Calculating memory, which monotonically rises throughout the binary's history, thus creates a discontinuity at the merger time which can not fully be remedied by techniques such as signal windowing.
Furthermore, frequency-domain waveforms are cut-off at \SI{20}{\hertz} which means that we can not perfectly restore the time domain waveform when we apply an inverse Fourier transform.
}.

Although precession effects are present at the population level, we do not expect these to cause substantially different memory estimates.
None of the individual binary black-hole mergers reported so far show significant signs of precession, though, GW190412 and GW190521 exhibit a mild preference for precession~\cite{GWTC1, GWTC2, GW190412}.
Additionally, we find in injection studies that precessing spins do not meaningfully change the signal-to-noise ratio of the memory part of the waveform.

Recent studies of GW190521 suggest that it may be a highly eccentric ($e\geq 0.2$) binary~\cite{IsobelGW190521, GayathriGW190521, GW190521Discovery}. 
Eccentric binaries may become an interesting target to measure memory in future generation detectors~\cite{FavataEccentricity}, however, since they are not yet firmly established, we assume all binaries to be circular for this work.

We measure memory by performing model comparison on each event by calculating a Bayes factor ($\BF_{\mathrm{mem}}$) for the presence of memory.
For a more in-depth discussion of these methods we refer to our previous work~\cite{Lasky2016, ThanksForTheMemory}.
The memory Bayes factor can be understood to be the fraction of evidences $Z_{\mathrm{osc + mem}}$ and $Z_{\mathrm{osc}}$ for the combined oscillatory plus memory waveform and oscillatory-only waveform, respectively
\begin{equation}
    \BF_{\mathrm{mem}} = Z_{\mathrm{osc + mem}} / Z_{\mathrm{osc}}\, .
\end{equation}
In order to calculate the Bayes factors in practise, we perform initial sampling runs with the \textsc{dynesty}~\cite{Dynesty} implementation within \bilby~\cite{2018Bilby, BilbyGWTC1} using the \imrx and \nrsur waveform models. 
We then use a modified version of those models which contains the memory contribution and obtain the Bayes factor for each event using importance sampling~\cite{HOM_REWEIGHTING, Thrane2019, ThanksForTheMemory}.
In order to calculate the memory contribution we adapt the \textsc{gwmemory} \textsc{Python} package~\cite{Talbot2018} to support \imrx and \nrsur and make additional modifications produce waveforms with a consistent length which is required for practical inference tasks.
Next, we obtain the memory Bayes factor by summing the memory weights $w_{\mathrm{mem}}$ over all $n_{\mathrm{post}}$ posterior samples $\theta_i$
\begin{equation}
    \mathrm{BF}_{\mathrm{mem}} = \frac{1}{n_{\mathrm{post}}}\sum_{i = 1}^{n_{\mathrm{post}}} w_{\mathrm{mem}}(\theta_i) \, .
\end{equation}
The weights are defined to be the ratios between the likelihoods $L$ of the two hypotheses
\begin{equation}
    w_{\mathrm{mem}}(\theta_i) = \frac{L_{\mathrm{mem+osc}}(\theta_i)}{L_{\mathrm{osc}}(\theta_i)}\, .
\end{equation}
Using importance sampling effectively suppresses stochastic sampling noise in the evidence calculation, making it more suitable than performing inference, e.g. using nested sampling~\cite{SkillingNestedSampling}, with both hypotheses separately and comparing the resulting evidence values~\cite{ThanksForTheMemory}.

Since including memory does not add any additional parameters to our problem, we do not need to consider effects due to increased prior volume.
We set our prior odds on the presence of memory to be 1, i.e. we give equal weight to memory either being present or not.
This means we could also interpret the memory Bayes factor as an odds.

\section{Results}

\begin{figure*}[ht]
    \centering
    \includegraphics[width=\textwidth]{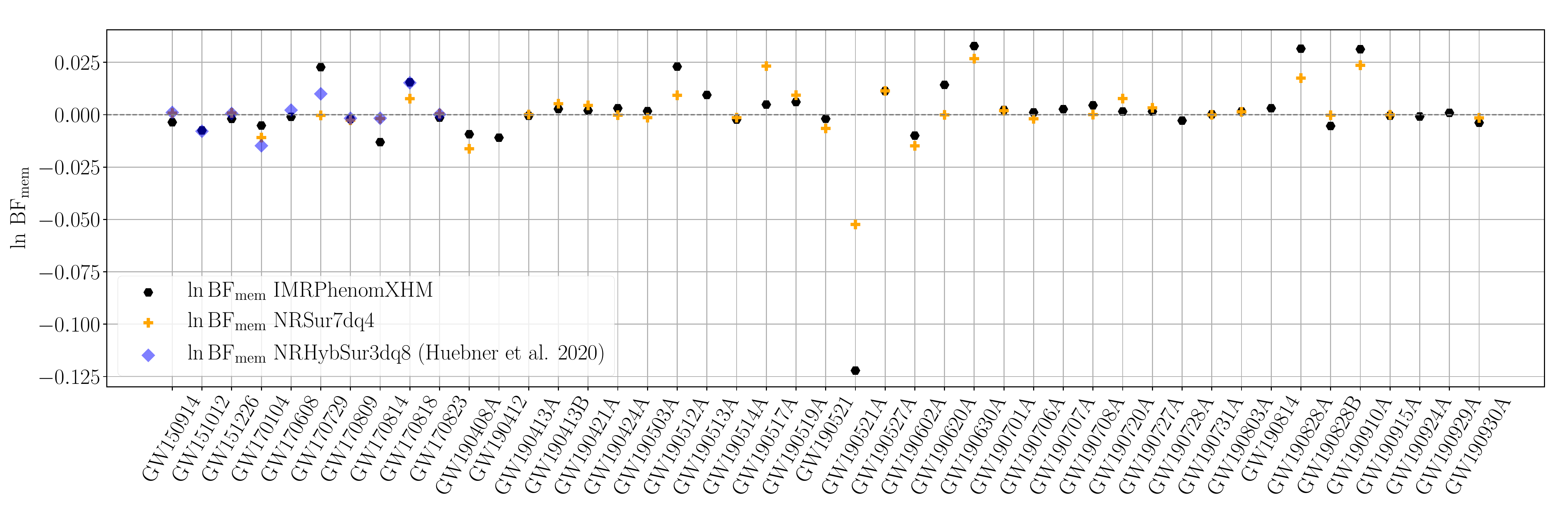}
    \caption{Memory Bayes factors obtained for the first and second gravitational-wave transient catalog using \imrx and \nrsur as fiducial waveform models, as well as the results from~\cite{ThanksForTheMemory} which used \nrhybsur. We abbreviate the last six digits of the regular event nomenclature with a simplified `A' or `B' to refer the first or second event detected on that date for readability. 
    Overall, there is no evidence for or against the memory hypothesis.
    }
    \label{fig:gwtm2}
\end{figure*}

As discussed in Ref~\cite{ThanksForTheMemory}, $\ln \mathrm{BF}_{\mathrm{mem}} > 8$ can be considered to be very strong evidence for the presence of gravitational-wave memory.
We present the new findings for the unambiguous 36 new binary-black hole observations plus GW190814 additionally to a re-analysis of the first ten binary-black hole observations in Fig. \ref{fig:gwtm2}.
We calculate a cumulative $\ln \mathrm{BF}_{\mathrm{mem}} = 0.025$ using \imrx and $\ln \mathrm{BF}_{\mathrm{mem}} = 0.049$ using \nrsur, which indicate that there is no strong evidence favouring or disfavouring the presence of memory in the signals.

We omit the analysis of GW190425 and GW190426\_152155, which are most likely to be a binary neutron star and a black hole-neutron star binary system, respectively, for two reasons.
First, low mass binaries produce far less memory within the LIGO band than heavier binary black holes, which makes them less useful for memory studies~\cite{Johnson2018}.
Second, neither \imrx nor \nrsur model neutron star physics, and we would thus need to implement and test another waveform model for very marginal benefit.
Additionally, GW190426\_152155 has a relatively high false alarm rate of 1.4 per year and thus might not be of astrophysical origin.
Since they are the other most likely events to not be of astrophysical origin we also exclude GW190719\_215514 and GW190909\_114149.
We are otherwise liberal and include all events that have been reported in the catalogs so far.
For the \nrsur runs, we exclude GW151012, GW190412, GW190814, GW190513\_205428, GW190707\_093326, GW190728\_064510, GW190924\_021846, and GW190929\_012149 since they show substantial posterior support for $q < 1/4$.

We visually verify that our parameter estimates are broadly consistent with what has been reported in~\cite{GWTC2}.
While we occasionally find minor differences, this is likely due to the fact that some of the runs in the catalog~\cite{GWTC2} were performed using a sampling frequency of \SI{512}{\hertz}, which implies that no physics beyond \SI{256}{\hertz} was included in the analysis. 
However, most events have some contributions at higher frequencies due to higher-order modes, which cause the posterior to shift.

While we do not find evidence for or against the presence of memory in any of the observed systems, some of the obtained Bayes factors stand out and some events deserve our attention.

\subsection{GW190521}
GW190521 is the highest mass event that has been reported so far~\cite{GW190521Discovery}.
Even though higher mass systems should in principle create more memory since they radiate off more energy in gravitational waves, their memory is shifted outside of the observable LIGO/Virgo band~\cite{Johnson2018}.
This is consistent with our finding that the measured memory Bayes factor indicates GW190521 to be uninformative about memory.
\subsection{GW190521\_074359 and similar events}

\begin{figure}[ht]
    \centering
    \includegraphics[width=\linewidth]{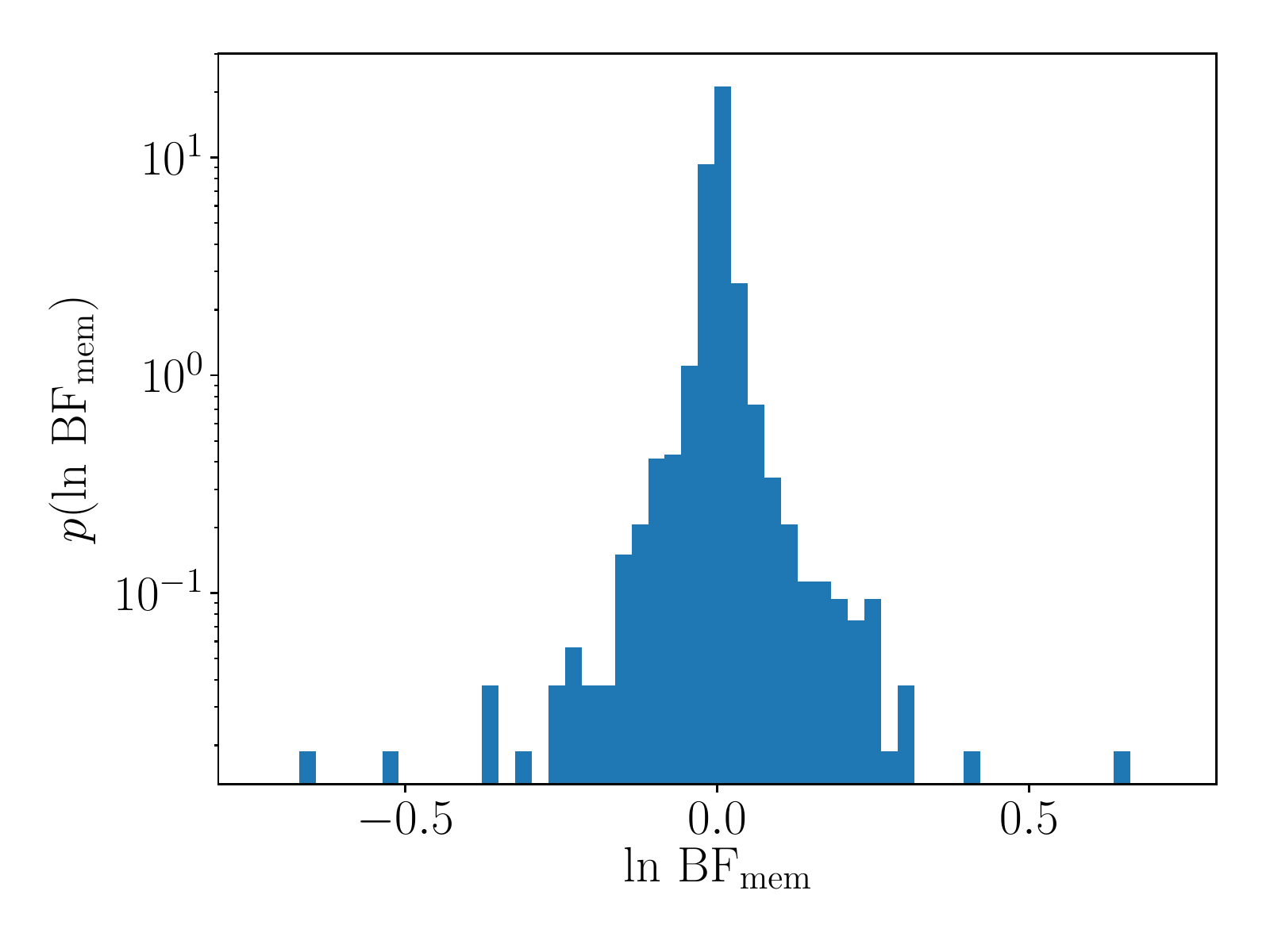}
    \caption{The distribution of Bayes factors for a simulated population of events from Ref.~\cite{ThanksForTheMemory} exhibits wide tails.
    For visualization, we do not show the two most extreme Bayes factors, which have values of $\ln \mathrm{BF}_{\mathrm{mem}} = 2.4$ and $4.0$.
    }
    \label{fig:ln_bf_distribution}
\end{figure}

GW190521\_074359 is a near equal mass binary with a total mass of around $75 M_{\odot}$ and is also one of the loudest events observed so far ($\mathrm{SNR} \approx 26$)~\cite{GWTC2}.
These properties generally point towards it being a favourable event with which to measure memory.
Despite this, we find $\ln \BF_{\mathrm{mem}} = -0.12$ using \imrx and $\ln \BF_{\mathrm{mem}} = -0.05$ using \nrsur, which is the lowest memory Bayes factor but the highest by absolute value for both waveform models. 
A negative $\ln \BF_{\mathrm{mem}}$ for any individual event is not concerning since they are expected to arise from noise fluctuations.
To show this, we re-examine the population study in Ref.~\cite{ThanksForTheMemory}.
While the population study is not perfectly comparable with the set of the actual measured events as the former is based on a point estimate of the inferred population of GWTC-1~\cite{LIGO_RATES_AND_POPULATIONS}, and used different waveform models, it can still provide us with a cross-check to see if the observed distribution of Bayes factors is sensible.
We find that out of 2000 simulated events, 28 have $\ln \BF_{\mathrm{mem}} < -0.12$  despite memory being present, which indicates that our measurement GW190521\_074359 is broadly consistent with our expectations.
Furthermore, as we show in Fig.~\ref{fig:ln_bf_distribution}, the distribution of Bayes factors has wide tails, which means that single outlying values are to be expected.

How is it then that the event that looks most likely to contain measurable memory returns the lowest memory Bayes factor in the catalog?
Events that are highly unfavourable to measure memory with will return $\ln \BF_{\mathrm{mem}} \approx 0$ as they can only be uninformative.
On the other hand, events like GW190521\_074359 are more informative about memory, but memory is still weak relative to detector noise. 
Hence, their signals are also prone to noise fluctuations that may randomly cancel out the memory contributions, which results in a negative log Bayes factor. 
This is only true for weak memory signals, though.
If the signal-to-noise ratio of the memory is greater than one, memory dominates over noise effects and it becomes much less likely that we measure a negative log Bayes factor due to noise fluctuations.
As Fig.~\ref{fig:ln_bf_distribution} demonstrates, the log Bayes factor of GW190521\_074359 is still in the regime in which we expect noise fluctuations to be able to change the overall sign of the result.

Other events with high absolute memory log Bayes factors (e.g. GW190630\_185205, GW190828\_065509, GW190910\_112807) follow a similar pattern to GW190521\_074359 in that they are relatively high signal-to-noise ratio events, close to equal mass, and have a total mass between $50-80M_{\odot}$.

\begin{figure}[ht]
\centering
\subfigure{\includegraphics[width=\linewidth]{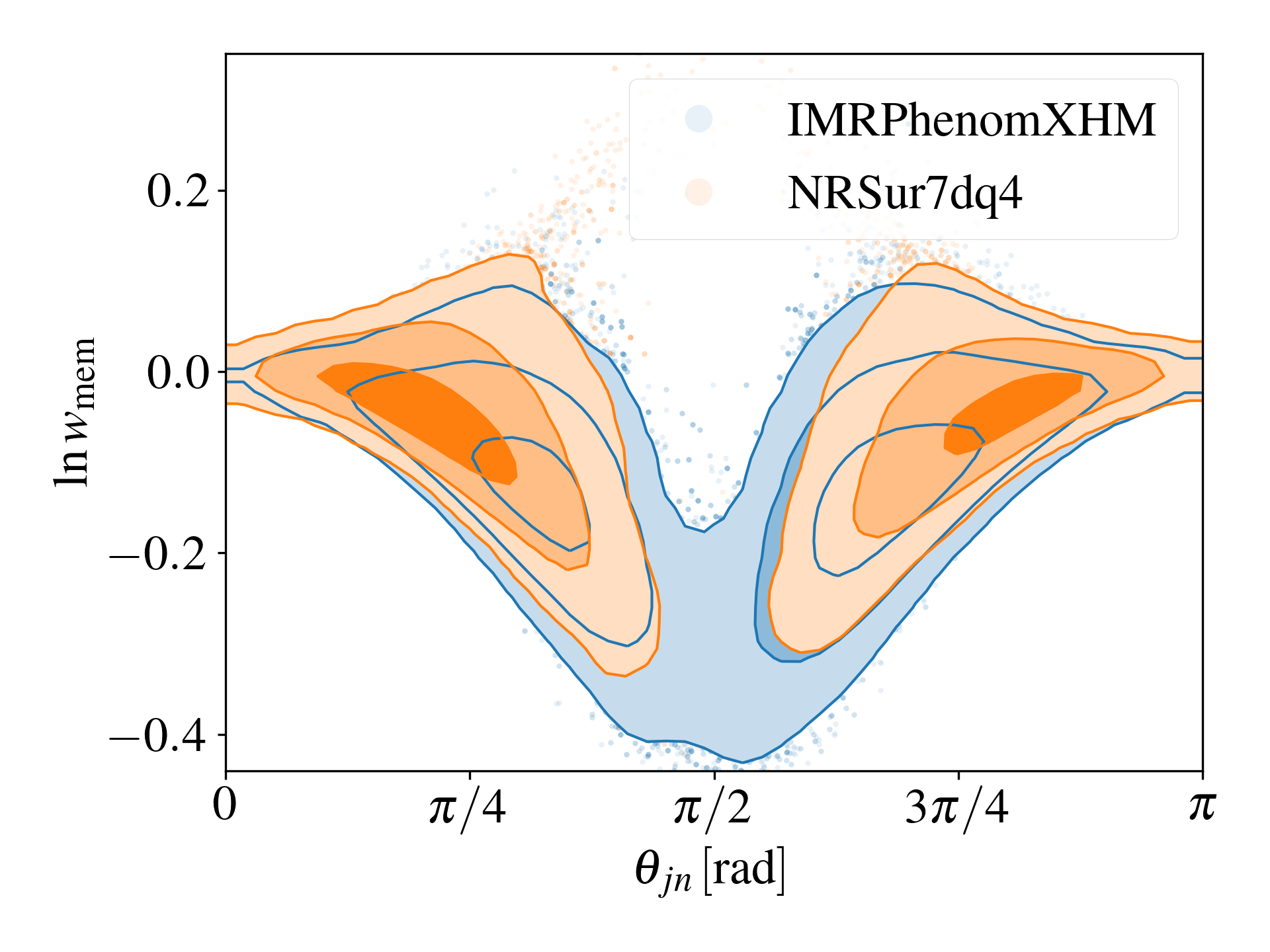}}
\subfigure{\includegraphics[width=\linewidth]{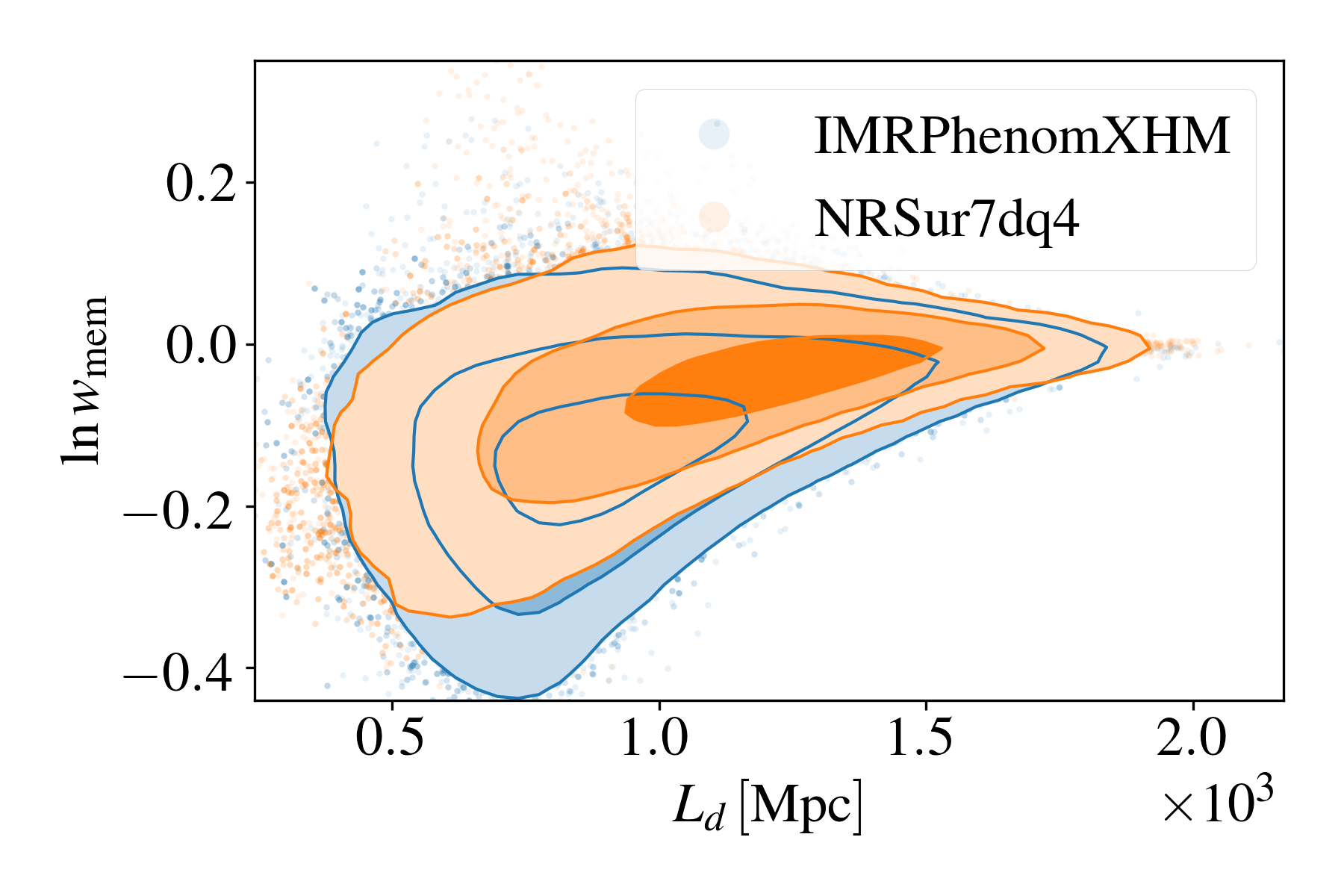}}
\caption{Two dimensional contour plots of the inferred posterior in terms of inclination and luminosity distance versus the calculated memory weights for \imrx (blue) and \nrsur (orange).
We show inclination ($\theta_{jn}$) in the top and luminosity distance ($d_L$) in the bottom subfigure.
The top subfigure demonstrates why the \imrx have samples with on average larger weights.
\nrsur samples are somewhat further constrained away from $\theta_{jn} = \pi/2$ which corresponds to an edge-on binary for which observed memory is maximal.
The bottom subfigure demonstrates that samples at closer distances have larger absolute log weights on average, corresponding to the fact that closer events are more informative than ones further away.
}
\label{fig:rorschach}
\end{figure}

While most of the differences between the Bayes factors from our two waveform models are minor, they do appreciably diverge for GW190521\_074359.
The difference is unlikely to be due to stochastic sampling noise as this is strongly suppressed in the importance sampling step ~\cite{ThanksForTheMemory}.
In order to understand this difference, we examine the posteriors of both \imrx and \nrsur.
In Fig.~\ref{fig:rorschach}, we display the posterior as a contour plot in terms of the obtained memory log weights $\ln w_{\mathrm{mem}}$, and the inclination angle $\theta_{jn}$ as well as the luminosity distance $d_L$.
The observed memory strain $h_{\mathrm{mem}}$ is highly sensitive to inclination angle $\theta_{jn}$ we are viewing the binary at
\begin{equation}
 h_{\mathrm{mem}}(\theta_{jn}) \propto \sin^2 \theta_{jn} (17 + \cos^2 \theta_{jn}) \, .   
\end{equation}
See e.g. Ref.~\cite{Favata2009} for a detailed derivation of this relation.
Thus, memory is most easily seen edge-on ($\theta_{jn} = \pi / 2$) as opposed to the oscillatory part which is preferably emitted face-on ($\theta_{jn} = 0,  \pi$).
In the posteriors, \nrsur has stronger support to be closer to face-on whereas \imrx shows support for GW190521\_074359 being an edge-on binary.
This leads to the \imrx weights obtaining larger absolute log weights.
The preference of \nrsur being closer to face-on conversely corresponds to a higher inferred luminosity distance than \imrx. 
Overall, these inferred differences in posteriors are expected due to systematic differences in the waveform models, e.g. \nrsur contains precession effects and all modes up to $(\ell, |m|) = (4, 4)$, whereas \imrx only has aligned spins and modes $(\ell, |m|) = (2, 2), (2, 1), (3, 3), (3, 2), (4, 4)$.

\subsection{Comparison with GWTC-1 analysis}

As part of our analysis we redo the analysis in Ref.~\cite{ThanksForTheMemory} for the events of the first two observing runs in which we originally used the hybridized surrogate model \nrhybsur~\cite{NRHYBSUR}.
We find that the difference between \nrhybsur and either \imrx and \nrsur to be in the same order of magnitude as differences between \imrx and \nrsur.
Again, this is most likely to be due to systematic differences in the waveform models, as well as, to some extent, sampling noise causing slight deviations.
We also note that stochastic sampling error scales with the square root of the number of events so an error of $10^{-2}$ per event would only scale up to an error of $\Delta (\ln \mathrm{BF})\approx 0.5$ for the $\mathcal{O}(2000)$ events required to reach $\ln \mathrm{BF} = 8$.

\section{Conclusion and Outlook}\label{sec:conclusion}
We implement the memory waveforms associated with two waveform models, \imrx and \nrsur, using the memory calculation method laid out in Refs.~\cite{Favata2010, Talbot2018} 
We perform Bayesian model comparison to search for the presence of memory in the data.
Using the \imrx (\nrsur) model we find a combined $\ln\BF_{\mathrm{mem}} = 0.025 \, (0.049)$ in the first and second gravitational-wave transient catalog.
This is consistent with our expectation that $\mathcal{O}(2000)$ events are required to reach $\ln\BF_{\mathrm{mem}} = 8$, which we consider to be very strong evidence~\cite{ThanksForTheMemory, BoersmaMemory}.
We find that differences in the Bayes factors for each event are likely due to systematic differences in the waveforms and to a lesser extent due to stochastic sampling noise.

We have shown that our approach outlined in our previous paper (Ref.~\cite{ThanksForTheMemory}) is scalable up to a large number of events, demonstrating the possibility to coherently search for memory in the future.
Given the rapid developments in the waveform community and innovations such as massively parallel Bayesian inference~\cite{PBilby}, we anticipate that more advanced waveform models can be used for inference in the future.
These waveform models may allow us to calculate higher-order and precessing effects at greater mass ratios and thus remove the need for using multiple waveform models to cover all observed events.

\section{Acknowledgements}\label{sec:acknowledgements}
We would like to thank Ethan Payne, Cecilio Garcia-Quiros, Colm Talbot, and Bernard Whiting for helpful discussions.
This work is supported through Australian Research Council (ARC) Centre of Excellence CE170100004. PDL is supported through ARC Future Fellowship FT160100112 and ARC Discovery Project DP180103155.
This material is based upon work supported by NSF’s LIGO Laboratory which is a major facility fully funded by the National Science Foundation.
This work was performed on the OzSTAR national facility at Swinburne University of Technology. The OzSTAR program receives funding in part from the Astronomy National Collaborative Research Infrastructure Strategy (NCRIS) allocation provided by the Australian Government.
This is LIGO Document No. DCC P2100125.
\bibliography{memory}

\end{document}